\documentclass[aps,pra,floatfix,twocolumn,showpacs,10pt,longbibliography]{revtex4-2}
\usepackage{amsmath}
\usepackage{url}
\usepackage[ruled,vlined]{algorithm2e}
\usepackage{algorithmic}
\usepackage{graphicx}
\usepackage{dcolumn}
\usepackage{bm}
\usepackage{amssymb}
\usepackage{rotating}
\usepackage[abs]{overpic}
\usepackage{xcolor}
\usepackage{tabularx}
\usepackage{floatrow}
\usepackage{subfig} 
\usepackage[justification=RaggedRight]{caption}
\usepackage{hyperref}
\hypersetup{colorlinks = true, citebordercolor={blue}, linkcolor={blue}, citecolor={blue}, urlcolor={blue}}
\begin{document}












\title{\color{black}Entangled \color{black} Phase of Simultaneous Fermion and Exciton Condensations Realized}

\author{LeeAnn M. Sager and David A. Mazziotti}

\email{damazz@uchicago.edu}

\affiliation{Department of Chemistry and The James Franck Institute, The University of Chicago, Chicago, IL 60637}%

\date{Submitted August 8, 2021\color{black}; Revised December 30, 2021\color{black}}




\begin{abstract}

Fermion-exciton condensates (FECs)---computationally- and theoretically-predicted states that simultaneously exhibit character of superconducting states and exciton condensates---are novel quantum states whose properties may involve a hybridization of superconductivity and the dissipationless flow of energy.  Here, we exploit prior investigations of superconducting states and exciton condensates on quantum devices to identify a tuneable quantum state preparation entangling the wavefunctions of the individual condensate states.  Utilizing this state preparation, we prepare a variety of fermion-exciton condensate states on \color{black} quantum computers---realizing strongly-correlated FEC states on current, noisy intermediate-scale quantum devices---\color{black}and verify the presence of the dual condensate via post-measurement analysis.  This \color{black} confirmation of the previously-predicted condensate state on quantum devices \color{black} as well as the form of its wavefunction motivates further theoretical and experimental exploration of the properties, applications, and stability of fermion-exciton condensates.

\end{abstract}



\maketitle

\section{Introduction}
It may be possible to create materials that conduct both electric current and exciton excitation energy through the realization a single quantum state that simultaneously demonstrates properties of two different condensates---one composed of ``Cooper'' (particle-particle) pairs and the other composed of excitons (particle-hole pairs) \cite{Sager_2019}.
Bose-Einstein condensation allows for multiple bosons aggregating in a single quantum state \cite{bose_einstein_1924,einstein_1924} at sufficiently low temperatures, resulting in the superfluidity of the constituent bosons \cite{london_1938,tisza_1947}.  A superconducting quantum phase is created upon the condensation of pairs of fermions into a single quantum state, which results in the frictionless flow of the constituent particle-particle pairs \cite{BCS1957,Anderson_2013}.  Significant theoretical and experimental investigation \cite{BCS1957, Blatt_SC, Anderson_2013, Drozdov_250, Ginzburg_1991, crabtree_2020, twist_2018, twist_2020_1, twist_2020_2,TSH2004, Fil_Shevchenko_Rev, Shiva, Kogar2017, LWT2017,london1950superfluids,FEYNMAN195517,leggett_1999,gorter_2011,guo_2020,hao_2020,Buzzi_2020,delpace_2021,Sager_2021} has centered on superconductors in an effort to determine a commercially-viable material supporting superconductivity at sufficiently high temperatures.  However, the relatively low energy of the Cooper pairs \cite{BCS1957,Anderson_2013} cause them to become unstable, reverting to traditional conductors above a critical temperature too low for commercial applications.

One avenue towards higher-temperature condensate phases has been an examination of condensations composed of particle-hole pairs (excitons) in a single quantum state, which can carry exciton excitation energy without resistance  \cite{Fil_Shevchenko_Rev, keldysh_2017}.  Excitons are more-tightly bound than Cooper pairs, meaning that the condensation of excitons can persist at higher temperatures than those at which superconductors form, although the natural recombination of particles and holes is a cause of experimental difficulties in creating stable, high-temperature, ground-state exciton condensates. As such, much recent literature has explored the characteristics of exciton condensation as well as established various methodologies for overcoming the problem of annihilation upon recombination \cite{KSE2004, TSH2004, Fil_Shevchenko_Rev, Shiva, Kogar2017, LWT2017, varsano_2017, fuhrer_hamilton_2016, Sager_2020}.  Specifically, exciton condensates have been observed in optical traps with polaritons \cite{KRK2006, BKY2014, DWS2002, BHS2007},  the electronic double layers of semiconductors \cite{SEP2000, KSE2002,KSE2004, TSH2004,NFE2012} and graphene \cite{LWT2017, Lee2016, Li2016}, and in systems composed of transition metal chalcogenides \cite{Debnath_2017,Kogar2017,Wang_2019,Liu_2020,Bretscher_2021,Jiang_2021}.


\begin{figure}[tbh!]
  \includegraphics[width=8.5cm]{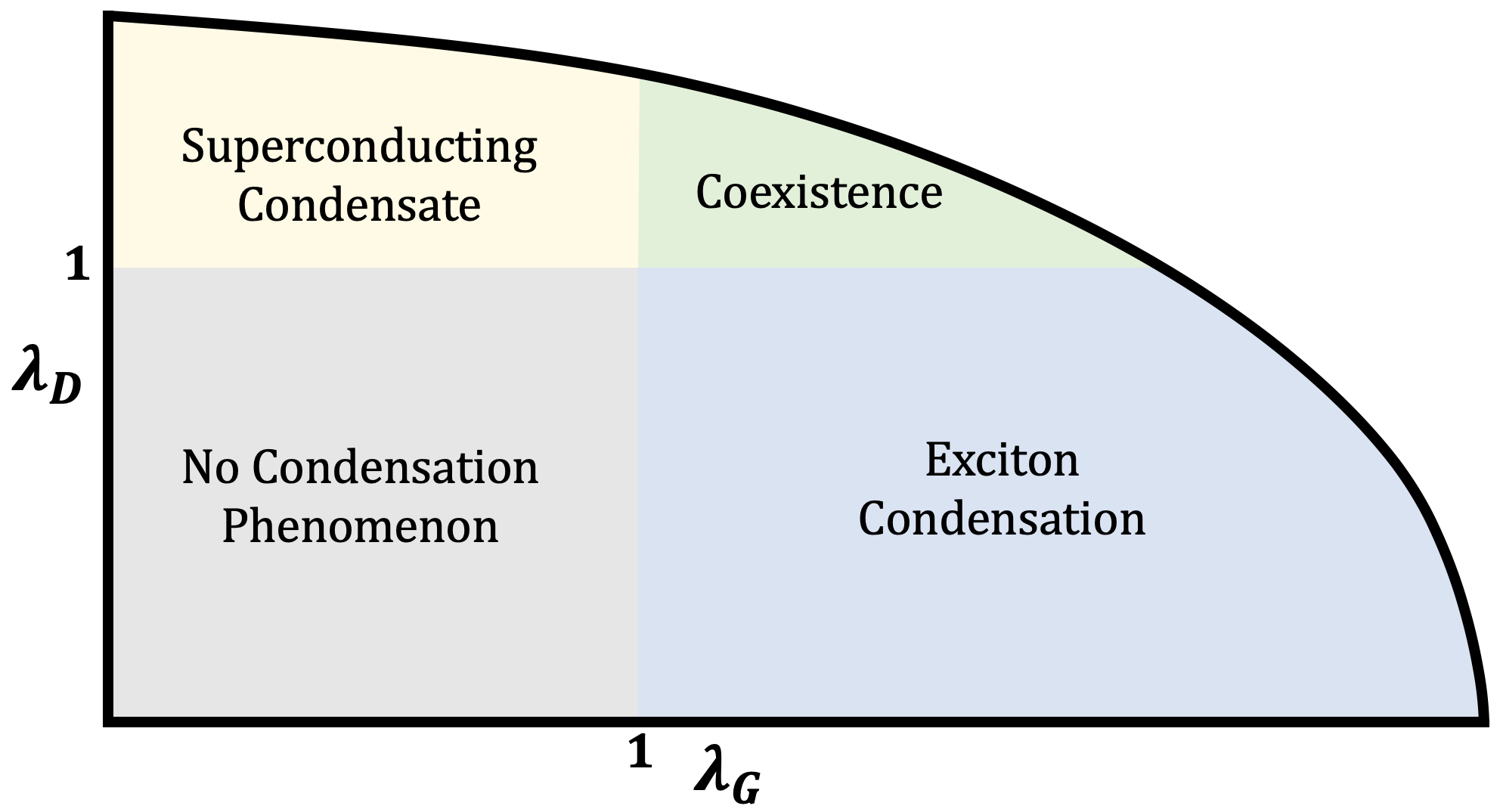}
  \caption{\label{fig:Tradeoff} A figure of the condensate phase diagram in the phase space of the signatures of particle-particle condensation, $\lambda_D$, and exciton condensation, $\lambda_G$,\color{black}---previously presented as Fig. 1 in Ref. \onlinecite{Sager_2019}---\color{black}is shown.}
\end{figure}

Here, we present \color{black} an entangled \color{black} quantum phase of matter in which a superconductor and an exciton condensate coexist in a single quantum state---a fermion-exciton condensate (FEC).  We leverage the ability of quantum computation to explore strongly correlated phenomena
\cite{wallraff_2004,koch_2007,goppl_2008,chow_2011,andersson_2015,schuster_2019} as well as prior investigation \cite{Sager_2020,Sager_2021} to prepare a variety of fermion-exciton condensate states on quantum device\color{black}s \color{black} via a tuneable quantum state preparation---verifying the presence of the condensate state through probing the signatures of particle-particle ($\lambda_D$) \cite{Y1962,S1965} and exciton ($\lambda_G$) \cite{GR1969,Shiva} condensations.  Our results not only confirm the existence of a new class of condensates, they verify the theoretical prediction of the form of the wavefunction of the FEC as well as the phase diagram of the states (see Fig. \ref{fig:Tradeoff}).  These results suggest that such condensates can potentially be prepared in physical systems such as twisted graphene bilayers in which forces favoring exciton condensation and superconducting, respectively, are in fierce competition.


Note that while both particle-particle and exciton condensates are known to exist in systems designed to use exciton condensates to mediate the creation of Cooper pairs at higher temperatures \cite{LTSK2012,SCKP2018}, this coexistence
of fermion pair and excitonic condensation occurs in two adjacent systems that interact with one another \cite{SCKP2018} 
instead of existing in a joint FEC state.

\section{Theory}
\subsection{Signatures of Condensation}
Condensation occurs when multiple bosons aggregate into a single quantum state \cite{bose_einstein_1924,einstein_1924} at temperatures below a certain critical temperature, resulting in the superfluidity of the constituent bosons \cite{london_1938,tisza_1947}.  In a condensation of particle-particle pairs, pairs of fermions condense into a single geminal, a two-fermion function directly analogous to a one-fermion orbital \cite{Y1962,C1963,S1965,RM2015,Srev_1999,shull_1959}, resulting in the frictionless flow of the particle-particle pairs \cite{BCS1957,Anderson_2013}.  As established by Yang \cite{Y1962} and Sasaki \cite{S1965}, a computational signature of such superconducting states---denoted as $\lambda_D$---is a large eigenvalue ($\lambda_D > 1$) of the particle-particle reduced density matrix (2-RDM) with elements given by
\begin{equation}
^{2} D_{k,l}^{i,j} = \langle \Psi | {\hat a}^{\dagger}_i {\hat a}^{\dagger}_j {\hat a}_l {\hat a}_k  | \Psi \rangle
\label{eq:D2}
\end{equation}
where $|\Psi\rangle$ is an $N$-fermion wavefunction and where $\hat{a}_i^\dagger$ and $\hat{a}_i$ are fermionic creation and annihilation operators for orbital $i$, respectively. This signature directly probes the presence and extent of non-classical (off-diagonal) long-range order \cite{RM2015}.

Exciton condensation, similarly, results from the condensation of particle-hole pairs (excitons) condensing into a single quantum state \cite{Fil_Shevchenko_Rev,keldysh_2017}.  A computational signature of exciton condensation---denoted as $\lambda_G$---is a large eigenvalue ($\lambda_G > 1$) of a modified version of the particle-hole reduced density matrix \cite{Shiva, GR1969, Kohn1970}, with elements given by
\begin{multline}
{}^{2}\Tilde{G}^{i,j}_{k,l}={}^{2}G^{i,j}_{k,l}-{}^{1}D^i_j{}^{1}D^l_k \\=\langle \Psi | {\hat a}^{\dagger}_i {\hat a}_j {\hat a}^{\dagger}_l{\hat a}_k  | \Psi \rangle-\langle\Psi|\hat{a}^\dagger_i\hat{a}_j|\Psi\rangle\langle\Psi|\hat{a}^\dagger_l\hat{a}_k|\Psi\rangle
\label{eq:modG2}
\end{multline}
where ${}^{1}D$ is the one-fermion reduced density matrix (1-RDM).  Note that this modification removes the extraneous large eigenvalue from a ground-state-to-ground-state transition.

See the Methods section of the Supplemental Material~\footnote{See Supplemental Material at (URL will be inserted by publisher) for state preparation, tomography of the reduced density matrices, and quantum device calibration data.} for specifics of how the signatures of superconductivity ($\lambda_D$) and exciton condensation ($\lambda_G$) are obtained from the result data of a given quantum state preparation.

\subsection{Fermion-Exciton Condensate}
A fermion-exciton condensate is a quantum state in which character of both particle-particle condensation and exciton condensation coexist \cite{Sager_2019}; thus, a fermion-exciton condensate exhibits simultaneous large eigenvalues of the particle-particle and modified particle-hole RDMs---i.e., $\lambda_D,\lambda_G > 1$. As we have previously theoretically established in the thermodynamic limit \cite{Sager_2019}, a fermion-exciton condensate should result from the entanglement of a wavefunction exhibiting superconductivity, $|\Psi_D\rangle$, with a wavefunction exhibiting exciton condensation, $|\Psi_G\rangle$), mathematically represented as
\begin{equation}
    |\Psi_{FEC}\rangle=\frac{1}{\sqrt{2 - |\Delta|}}\left(|\Psi_D\rangle - {\rm sgn}(\Delta) |\Psi_G\rangle \right),
\label{eq:FEC}
\end{equation}
where $\Delta = 2\langle \Psi_D | \Psi_G \rangle$ \cite{Sager_2019}.

\begin{figure*}[tbh!]
        \centering
        \includegraphics[width=16cm]{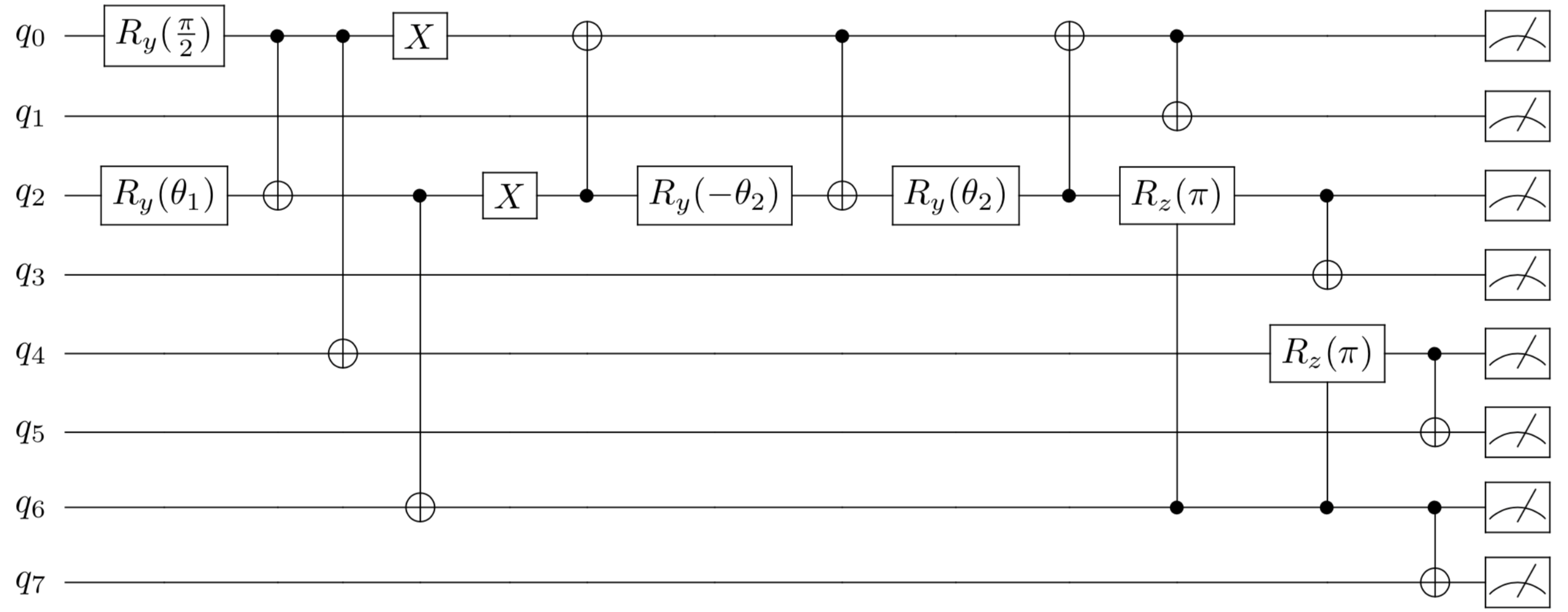}
        \caption{\label{fig:N8_dual} A schematic demonstrating the fermionic quantum state preparation that yields an entanglement of the non-zero elements of the separate particle-particle and particle-hole condensates \cite{Sager_2020,Sager_2021}, where $R_y$ and $R_z$ represent rotations about the $y$ and $z$ axes of the Bloch sphere and where two-qubit gates are represented such that the control qubit is specified by a dot connected to the target qubit, which is specified by the appropriate gate.    The wavefunction that results from this state preparation is given in the Supplemental Material.  Note that the condensate character---and hence the signatures of condensation $\lambda_G,\lambda_D$---are varied by scanning over input angles ($\theta_1,\theta_2$).}
\end{figure*}

From our previous work \cite{Sager_2019}, we note that a fermion-exciton condensate state is accessible in systems as small as four fermions ($N=4$) in eight orbitals ($r=8$), and from our investigations of condensate behavior on quantum devices, wavefunctions demonstrating maximal particle-particle condensation \cite{Sager_2021} and maximal exciton condensation \cite{Sager_2020}, individually, have been identified, prepared, and probed on quantum devices for $N=4,r=8$ systems.  Using the forms of these wavefunctions, we construct a state preparation that allows for the entanglement of the non-zero elements of the separate condensates, which is shown in Fig. \ref{fig:N8_dual}.  The input angles ($\theta_1,\theta_2$) are then optimized to generate a fermion-exciton condensate with character of each (i.e., a dual maximization of $\lambda_D$ and $\lambda_G$).

See the Methods section of the Supplemental Material for details of state preparations using both the bosonic representation of a qubit---in which each qubit is interpreted as a two-fermion geminal---and the fermionic representation of a qubit---in which each qubit is interpreted as a one-fermion orbital---as well as the optimization procedure for the input angles.  Also note that---as in Ref. \onlinecite{Sager_2021}---in our fermionic preparation, the pairing of qubits causes the usual difference between fermion and qubit statistics to disappear and hence allows for the direct representation of electron pairs on the quantum computer.

\section{Results}
\begin{figure*}[tbh!]
    \centering
    \subfloat[Santiago]{\label{fig:San}\includegraphics[width=8.5cm]{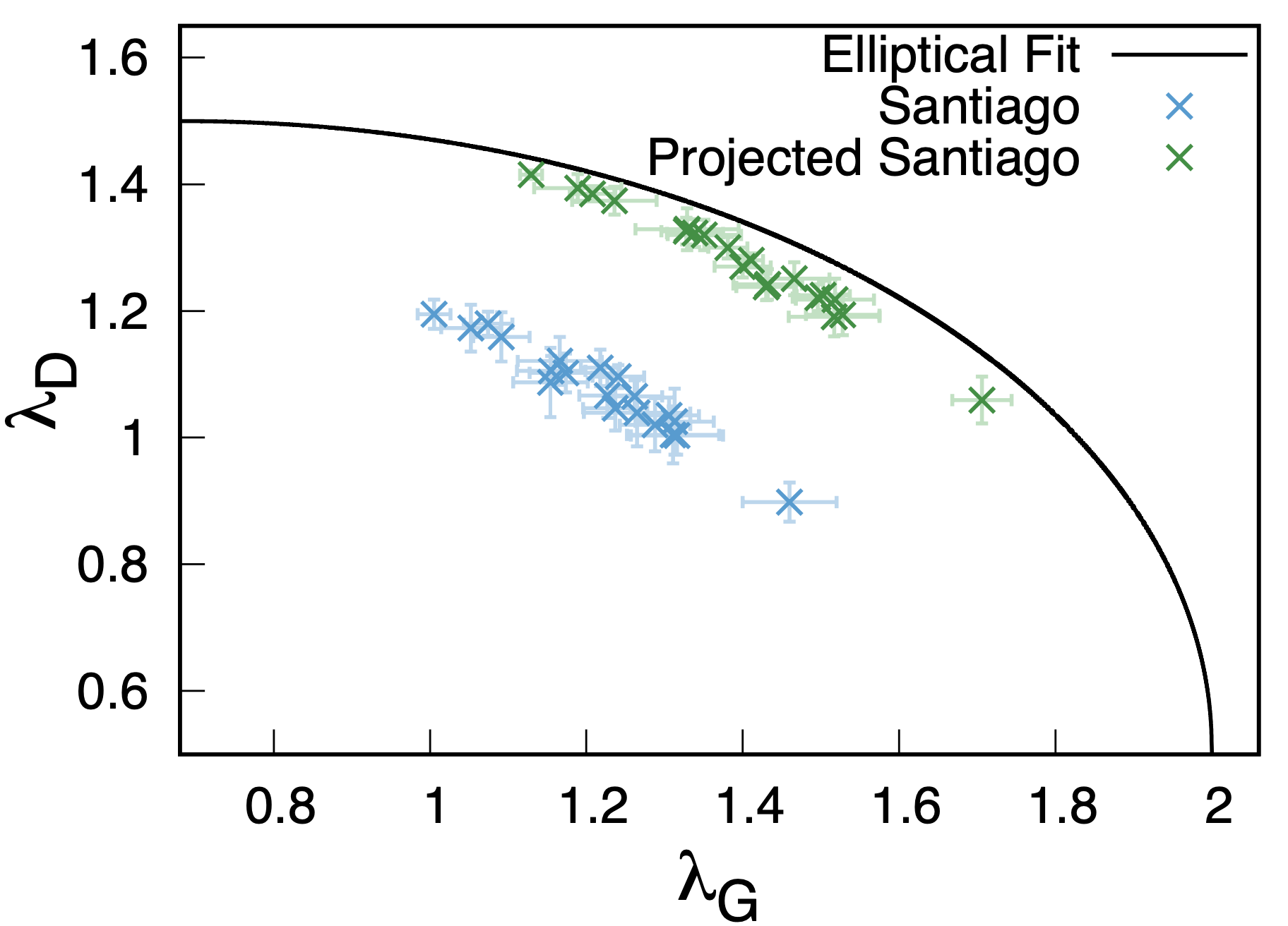}}
    \subfloat[Melbourne]{\label{fig:Mel}\includegraphics[width=8.5cm]{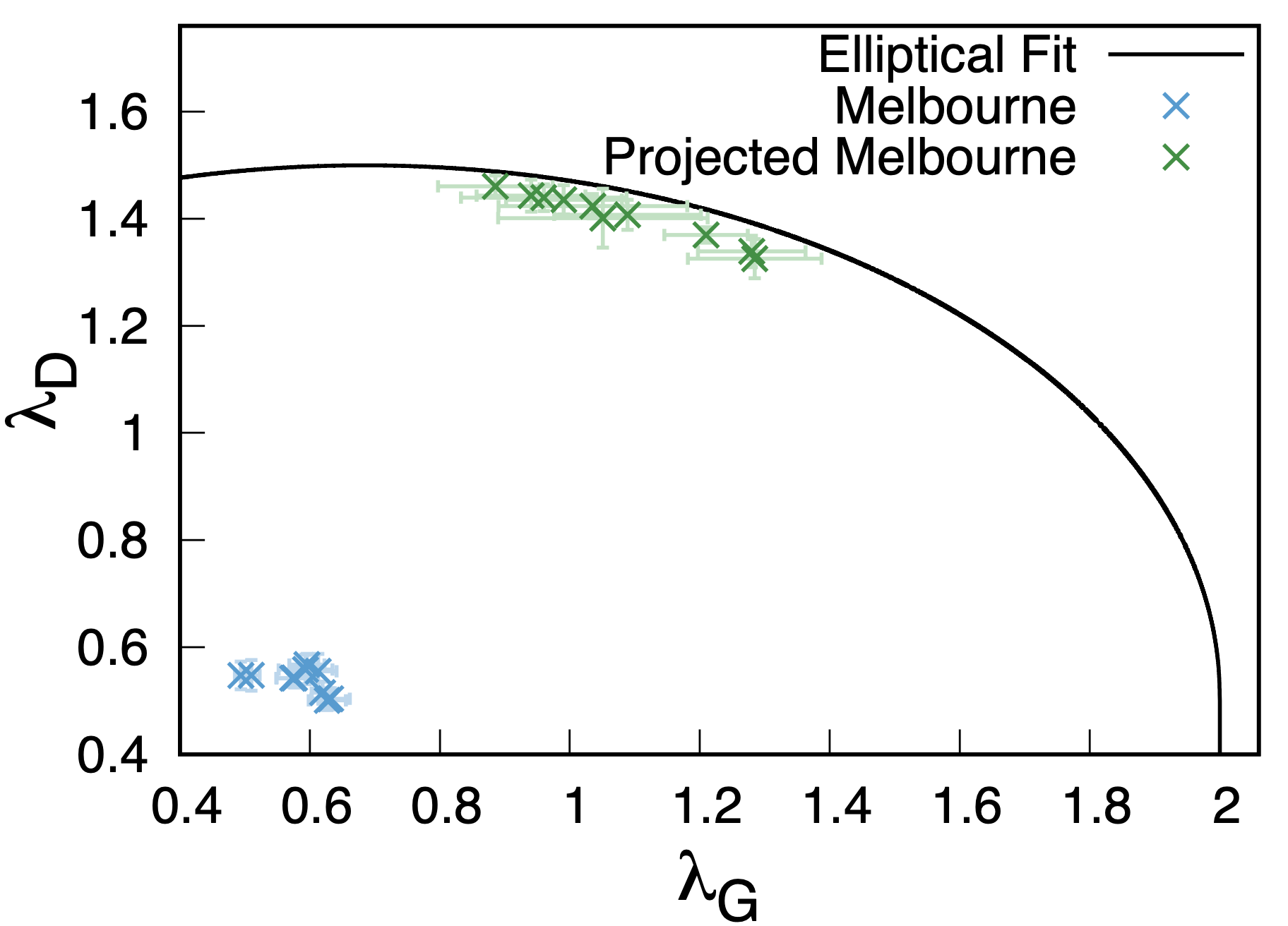}}
    \caption{\label{fig:san+mel}The eigenvalues of the ${}^2D$ and ${}^2\Tilde{G}$ matrices ($\lambda_D$ and $\lambda_G$, respectively) for various states prepared on IBM Quantum's (a) Santiago \cite{santiago} and (b) Melbourne \cite{melbourne} quantum computers before and after error mitigation via projection are plotted against the elliptical fit  \cite{Sager_2019} obtained from the unconstrained scan of $\lambda_D$ versus $\lambda_G$.  Note that the average value and standard deviation of ten trials per state preparation are shown.}
\end{figure*}

Using the bosonic state preparations with input angles that span the region exhibiting dual condensate character, we \color{black} prepare \color{black} a fermion-exciton condensate on a five-qubit quantum device \cite{santiago}.  As can be seen in Fig. \ref{fig:San}, with the blue x's representing device data before any error mitigation, various input angles yield quantum states with the signatures of both superconductor ($\lambda_D$) and exciton condensate ($\lambda_G$) character simultaneously exceeding the Pauli-like bound of one.  Moreover, as statistical analysis was conducted with the average and standard deviation determined from a sample size of ten trials per state preparation, these large eigenvalues are not spurious; rather, they are statistically-significant within one standard deviation. Further, for several of these unmitigated, experimental fermion-exciton condensates, the signature of condensation persists within two standard deviations.  (See the Supplemental Material for the standard deviation ranges of the signatures of condensation.)  \   Note that divergence from the maximal dual condensate character predicted for these input angles (i.e., the degree to which the reported data deviates from points along the elliptical fit) is likely due to preparation and readout errors on this noisy intermediate-scale quantum (NISQ) computer \cite{pakin_coles_2019,qiskit_errors,preskill_2018,zhang_2019}. (See the Supplemental Material for device error specifications.)

As is shown in Fig. \ref{fig:Mel}, using the fermionic state preparation on a noisier, fifteen-qubit quantum device \cite{melbourne} fails to realize a fermion-exciton condensate before error mitigation.  Further, as shown in Table \ref{tab:CompError_Sims} which presents the $\lambda_D$ and $\lambda_G$ values for a range of preparations simulated using four noise models that simulate errors consistent with real-world quantum device backends, this fermionic preparation would likely not yield fermion-exciton condensates on even the newer and less-error-prone Montreal and Mumbai quantum computers.  Likely, the four additional two-qubit, CNOT gates introduced into the fermionic state preparation---relative to the bosonic state preparation---introduce sufficient error to the quantum state such that the degree that condensate character is decreased or lost altogether.  This is further evidenced by both simulated Montreal and Mumbai being capable of demonstrating dual condensate behavior indicative of a fermion-exciton condensate for the bosonic state preparation and by simulated Melbourne demonstrating higher signatures of condensation for the bosonic preparation relative to the fermionic preparation.

In order to use NISQ devices to better-model these fermion-exciton condensate phases, we introduce an error mitigation scheme.  As can be seen in the Methods section of the Supplemental Material, the state preparations should yield quantum states with only six of the qubit basis states contributing to the overall wavefunction.  Any contribution from states other than these six basis states are unexpected and are directly caused by error on the quantum devices (or simulators) employed.  As such, we perform an error mitigation technique in which we project contributions from the qubit basis states that are not expected to contribute to zero and renormalize the resultant wavefunction.  As can be seen in Fig. \ref{fig:san+mel} and Table \ref{tab:CompError_Sims}, using this error mitigation technique improves results from both the fermionic and bosonic preparations.  Specifically---as can be readily observed from the green x's representing the mitigated, projected results in Fig. \ref{fig:san+mel}---, this projection technique leads to values approximating the ideal dual existence of excitonic and fermionic behavior along the elliptical fit, allowing us to prepare and probe ideal fermion-exciton condensates despite significant amounts of error on the NISQ quantum devices.

One interesting aside is that---for both the raw and projected data---the trade-off between character of a superconductor and that of an exciton condensate first noted in Ref. \onlinecite{Sager_2019} is also observed here.  This trade-off appears to be elliptic in nature---consistent with the convex nature of the 2-RDMs when projected onto two dimensions \cite{schwerdtfeger_mazziotti_2009,gidofalvi_mazziotti_2006,zauner_2016}---even for the noisy, non-mitigated Santiago results, and nearly the exact elliptic fit established in Ref. \onlinecite{Sager_2019} is observed when the contributions from the components that should not contribute are projected to zero. (See Ref. \onlinecite{Sager_2019} for additional details.)  This trade-off is significant as it precludes a fermion-exciton condensate with maximal condensate character of both particle-particle and exciton condensations.  However, as the trade-off is elliptic in nature and as the maximal $\lambda_D$ and $\lambda_D$ values increase with system size ($N$), the lengths of the major and minor axes of the elliptical fit will increase as the size of the system is increased, causing the trade-off to become less and less stark.

\begin{table}[]
\begin{tabular}{|cc|cccc|}
\hline
\textbf{Computer}  & \textbf{Quantum}          & \multicolumn{2}{c}{\textbf{Full}}        & \multicolumn{2}{c|}{\textbf{Projected}}   \\
          &     \textbf{Volume}        & $\mathbf{\lambda_G}$ & $\mathbf{\lambda_D}$ & $\mathbf{\lambda_G}$ & $\mathbf{\lambda_D}$ \\ \hline
\textbf{Fermionic} & \textbf{Preparation} &             &             &             &             \\ \hline
Melbourne & 8           & 0.804       & 0.994       & 1.191       & 1.352       \\
          &             & 0.921       & 0.597       & 1.387       & 1.303       \\
          &             & 0.934       & 0.581       & 1.440       & 1.271       \\
          &             & 0.910       & 0.567       & 1.461       & 1.261       \\ \hline
Montreal  & 128         & 0.918       & 1.134       & 1.169       & 1.329       \\
          &             & 1.193       & 0.868       & 1.472       & 1.255       \\
          &             & 1.241       & 0.849       & 1.527       & 1.215       \\
          &             & 1.267       & 0.824       & 1.580       & 1.176       \\ \hline
Mumbai    & 128         & 0.866       & 1.051       & 1.217       & 1.334       \\
          &             & 0.928       & 0.697       & 1.360       & 1.319       \\
          &             & 0.925       & 0.636       & 1.435       & 1.278       \\
          &             & 1.113       & 0.721       & 1.537       & 1.212       \\ \hline
\textbf{Bosonized} & \textbf{Preparation} &             &             &             &             \\ \hline
Melbourne & 8           & 0.875       & 1.256       & 1.130       & 1.377       \\
          &             & 1.013       & 0.919       & 1.328       & 1.334       \\
          &             & 1.077       & 0.906       & 1.409       & 1.289       \\
          &             & 1.094       & 0.914       & 1.422       & 1.281       \\ \hline
Montreal  & 128         & 1.126       & 1.281       & 1.244       & 1.310       \\
          &             & 1.306       & 1.107       & 1.454       & 1.265       \\
          &             & 1.400       & 1.066       & 1.547       & 1.205       \\
          &             & 1.406       & 1.041       & 1.567       & 1.188       \\ \hline
Mumbai    & 128         & 0.934       & 1.295       & 1.138       & 1.364       \\
          &             & 1.160       & 1.022       & 1.408       & 1.293       \\
          &             & 1.080       & 0.965       & 1.376       & 1.311       \\
          &             & 1.233       & 0.990       & 1.497       & 1.238       \\ \hline
Santiago  & 32          & 1.169       & 1.278       & 1.264       & 1.304       \\
          &             & 1.344       & 1.130       & 1.465       & 1.260       \\
          &             & 1.390       & 1.100       & 1.517       & 1.225       \\
          &             & 1.460	      & 1.060	    & 1.589	      & 1.174 \\\hline
\end{tabular}
\caption{\label{tab:CompError_Sims} Table of eigenvalues for the ${}^2\Tilde{G}$ ($\lambda_G$) and ${}^2D$ ($\lambda_D$) matrices obtained from noise model simulating errors from real-world quantum computers both before (full) and after (projected) error mitigation via projection of appropriate components to zero.}
\end{table}

\section{Conclusions}
In this paper, we prepare a fermion-exciton condensate---a single quantum state demonstrating both superconductivity and exciton condensation---on a quantum device.  This both realizes a \color{black} highly-correlated \color{black} state of matter \color{black} on a noisy intermediate-scale quantum device \color{black} and verifies the theoretical hypothesis from Ref. \onlinecite{Sager_2019} that such a state can be generated by entangling wavefunctions that separately exhibit particle-particle and exciton condensation. \color{black} Further, the error mitigation technique introduced leads to signatures of fermion-pair ($\lambda_D$) and exciton condensation ($\lambda_G$) approaching the ideal dual existence of excitonic and fermionic behavior along the elliptical fit on NISQ devices, allowing for better-modelling of these highly-correlated dual condensate phases on even extremely-noisy devices.\color{black}

\color{black} An \color{black} experimental system that may result in such an entanglement may be a bilayer system---as bilayers are often known to exhibit exciton condensation---in which the geometric orientation of the layers such as the twist angles are optimized to generate competition between forces favoring an exciton condensate and a superconductor.  It may also be possible to consider bilayers in which each layer is composed of a traditional superconductor, which can demonstrate particle-particle condensation.  These systems should be studied both computationally and experimentally as there are many open questions regarding the formation, properties, application, and stability of fermion-exciton condensates.  The possibility of a hybridization of the properties of superconductors with those of an exciton condensate definitely motivate further examination of this new state of matter.


\begin{acknowledgments}
\textit{Acknowledgments}: D.A.M. gratefully acknowledges the Department of Energy, Office of Basic Energy Sciences, Grant DE-SC0019215 and the U.S. National Science Foundation Grants No. CHE-2035876, No. DMR-2037783,  No. CHE-1565638\color{black}, and DGE-1746045\color{black}.
\end{acknowledgments}

{\it Data availability.} Data will be made available upon reasonable request.

{\it Code availability.} Code will be made available on a public Github repository upon publication.

\bibliography{references,references_2}

\end{document}